# RESEARCH AND INNOVATIVE DESIGN OF A ZERO-EMISSIONS VEHICLE BY MULTIDISCIPLINARY STUDENT TEAMS IN MULTI-YEARS

**Raymond K. Yee**, Professor
**Tai-Ran Hsu**, Professor and Chair, ASME Fellow
**Thuy T. Le**, Professor
San José State University
San José, California, USA

*Abstract*

This paper presents a unique learning and research experience for students from mechanical and electrical engineering majors in a course on senior design projects involving research & development, design and production of a proof-of-concept electric vehicle, the ZEM (Zero EMissions) vehicle. The ZEM vehicle combined positive aspects and latest technologies in electric vehicle design, solar/electric power conversions, and ergonomic human power into one affordable and environmentally sustainable vehicle for urban transportation. The 43 mechanical and 10 electrical engineering majors plus 7 students from business participated in this multidisciplinary project spanned over two academic years. The students involved in this multi-year endeavor gained valuable experiences in real-world working environment with multifunctional and multi-year sub-groups. The success of this new attempt in conducting senior design projects classes have set a model for faculty members in the authors' university in conducting similar courses.

I. INTRODUCTION

Authors of this paper have been instructors of a senior project course in their respective departments for many years. Previous projects, like those from other engineering schools, were focused on design projects on single-function devices or systems such as robots for specific tasks, human-power vehicles for annual ASME sponsored competitions and electric systems for certain devices. Most these design projects lack long lasting values and impacts to the community. The authors were motivated to experiment with a new approach in conducting this course by challenging their students to be involved in projects that would have potential benefits to global community in eco-environment and socioeconomic well being, and also acquire R&D experience from a real-world working environment with multifunctional and multi-year sub-groups. Consequently, they selected a pilot project of innovative design of a hybrid human/solar/electric-powered vehicle that would be practical for modern urban transportation in their senior design projects class in two academic years from 2006 to 2008.

There were 16 ME and 3 electrical engineering (EE) students signing up in an innovative design of a 4-wheel electric vehicle with the size that is compatible to that of neighborhood electric vehicles in the marketplace, and the vehicle could be driven by a similar hybrid human pedaling and electric power but with solar photovoltaic (PV) cells to supply additional power to charge the batteries. This vehicle is regarded as the ZEM vehicle. All authors acted as faculty supervisors of this project. In this first academic year, the student teams accomplished the design of a proof-of-concept prototype ZEM vehicle ready for construction. A team of 7 students from the College of Business at the authors' university used the ZEM project as its case study project for a course on business entrepreneurship in 2007. Student teams with expanded memberships of 27 ME and 7 EE students enrolled in the next academic year to undertake the construction of the prototype ZEM vehicle under the supervision of the second and third authors. This major task was accomplished in December 2008 with able assistance of technical staff of the College of Engineering in fault-proof test runs during the summer of that year. The design of ZEM vehicle won the First Prize in the 2007 National Idea-to-Product Competition for Engineering Projects in Community Services and Social Entrepreneurship at Princeton University with a cash award of $15,000.

II. THE ZEM PROJECT

There have been clear messages sent to engineering educators by prominent business leaders in the country that they desire engineering graduates to be proficient in: (1) communication, (2) team work experience, and (3) knowledge and experience in multidiscipline. These desired skills are documented in several special reports by the National Science Foundation, American Society for Engineering Education and the ABET [1-4]. Engineering educators have been working relentlessly to cultivate their students' acquiring such skills in addition to the professional knowledge and skill in their specific disciplines [5-7].

The ZEM vehicle, designed for 2 riders, is intended for both energy-efficient urban commuting and for light utility services and small business deliveries. The motor of the ZEM vehicle is driven by batteries that can be charged by household 110-VAC electric outlets, and by solar energy harnessed by the collection panels that are integrated into the exterior surface of the vehicle. The vehicle can be driven at low speed by human pedaling in inner city streets with congested pedestrian and vehicle traffic. The electric motor can power the vehicle to run at a higher speed in other parts of the city. Solar energy reduces current load from the batteries during the operations of the vehicle and charges the batteries when the vehicle parks in open parking lots. The ZEM vehicle is suitable for urban commuting, errand runners, small business deliveries and shuttle services between designated locations.

III. DESIGN AND DEVELOPMENT PROCESS

The ZEM vehicle project intends to produce a low cost, energy efficient vehicle that harnesses three types of energy: human pedaling, electro-chemical and photovoltaic. The design, research & development, and production of the "proof-of-concept" prototype ZEM vehicle turned out to be a major challenge to the students and the supervising faculty; there are several original ideas and concepts involved in the development process. Students conducted research and development on several key subsystems to warrant the success of the project:

A. Conduct research on suitability of batteries for the ZEM vehicle. Batteries are important design parameters for electric vehicle design; many batteries used in electric vehicles are heavy in weight and they become "dead load" to the vehicle after having consumed all the electric power they have stored. Two types of batteries were considered: the lead-acid batteries and silicone batteries. The lead acid batteries are low cost but have low energy density. The silicone batteries are relatively new on the market. They use dry silicate as electrolytes and have high power density suitable for vehicles, and they are also environmentally friendly. However, these batteries are high in cost. An extensive research study was spent by the student team to determine which of these two types of batteries would be suitable for the ZEM vehicle. Figure 1 shows the schematic of their study and their findings were presented in Table 1.

B. Research on solar-electric power conversion and transmission for ZEM vehicle. The teams of students of ME and EE majors soon realized that there was no precedence on the design in solar-electric energy conversion for vehicles in open literature. Several experimentations were performed to evaluate the feasibility of this subsystem design. An energy system experimentation was carried out. The objective was to quantitatively demonstrate the functionality and interaction between power produced by the solar PV cells and the power consuming systems for the ZEM vehicle.

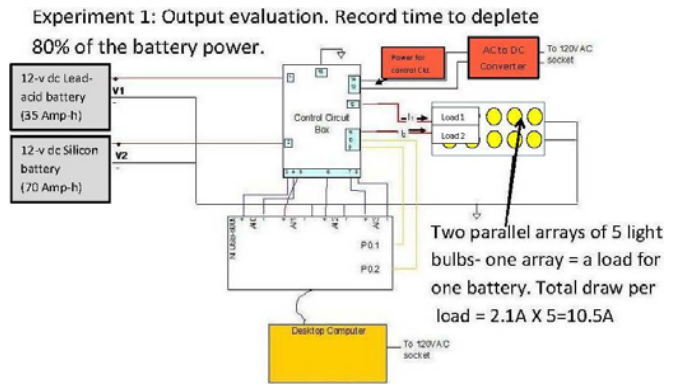

Fig. 1 Schematic of Battery Experiment

Table 1 EXPERIMENTAL FINDINGS

| Characteristic | Lead Acid | Silicone | Source |
|---|---|---|---|
| Battery Manufacturer (model) | Stinger (SPV35, AGM) | Clean Power Tech. Inc. (12V 17Ahr) | |
| Electrolyte (Environmental merit) | Sulfuric Acid (Acid mist, polluting) | Silicone compound (nonpolluting) | Ref. [8] |
| Theoretical Energy Density (Wh/Kg) | 35-44 | 45-52 | Ref. [8] |
| Experimental Energy Density (Wh/Kg) | 23.4 | 35.8 | ZEM Energy Experiment |
| Theoretical Amp-hours (nominal) | 35 (Discharging at 1.75A) | 70 (unknown discharge rate) | Respective Manufacturer |
| Experimental Effective Am-hours (at 12A, 100 to 20% SoC) | 21 | 43.8 | ZEM Energy Experiment |
| Charging Time | 4-8 hours | 2-4 hours | Ref. [8] |
| Memory Effects | @ low volt | None | Ref. [8] |

This experiment was carried out by a special facility on a roof-top balcony of the engineering building. It involved the selected high-output silicone batteries for the ZEM vehicle, PV charge controller, PC & power supply, and data acquisition system. In this experiment, students evaluated the solar PV system performance. Experiment was performed on an eight-hour day in March 2008 and the data was plotted against the theoretical prediction of available solar energy at this particular location. Table 2 shows the theoretical predictions and experimental measurements. The experimental measured data is averaged 26% lower than the predicted value in the solar PV output power. The main source of error was most likely from the environment of the experimental site which was surrounded by trees and buildings. The secondary source of error may have to do with the surface polarization effect [9] inherent in the SunPower's high efficiency solar cells design, although only a minor reduction in output power (2 to 5%) is expected in this application.

TABLE2 SOLAR PV EXPERIMENATAL FINDINGS

|  | Solrad.xls Predictions | Experimental Measurements | % Difference |
|---|---|---|---|
| Time Period of Measured Solar-Battery Charging | 0844a to 1624p | 0844a to 1625p | - |
| Average Global Solar Radiation (W/m^2) Bird Model - Direct Radiation | 439 | - | - |
| Maximum Global Solar Radiation (W/m^2) | 582 | - | - |
| Time of Max Radiation | 1:19PM | 12:25PM | - |
| Continuous PV Array Output (Watts) | 360 | 265 | 26 |
| Including 5% Decrease due to SPE (W) | 342 | 252 | 23 |
| Including 30% Decrease due to SPE (W) | 252 | 186 | 5 |
| Maximum PV Array Output (Watts) | 478 | 347 | 27 |
| Including 5% Decrease due to SPE (W) | 454 | 330 | 24 |
| Including 30% Decrease due to SPE (W) | 335 | 243 | 4 |

Another energy experiment was conducted to simulate the full energy system. This experiment provided feedback of a simulated system and data prior to actual full energy system integration on the ZEM vehicle. Figure 2 shows schematic of this energy system experimental setup. The above experiments allowed the students to not only become familiar with solar-electric conversion system for vehicles, but also the proper connections of various components and their integration into the ZEM vehicle.

C. Conduct research to prepare ZEM components for manufacturing/assembly, the ME student teams were working diligently on various tasks in the mechanical aspects of the ZEM project. ZEM vehicle consists of innovative design of several major mechanical systems including vehicle frame and chassis, power transmission system, and suspension system. Figure 4 shows the schematic of the electromechanical power transmission system used in the ZEM vehicle. Students developed a dual pedaling system for the ZEM vehicle which employs one common drive axle that can accept two separate power inputs; one from human pedaling and the other from an electric motor. The front wheel drive axle consists of two power transmitting sprockets which are integrated into the differential housing of the vehicle. One of the sprockets connects to the electrical motor and the other sprocket connects to the pedal input shaft for human pedaling as shown in Figure 3 below.

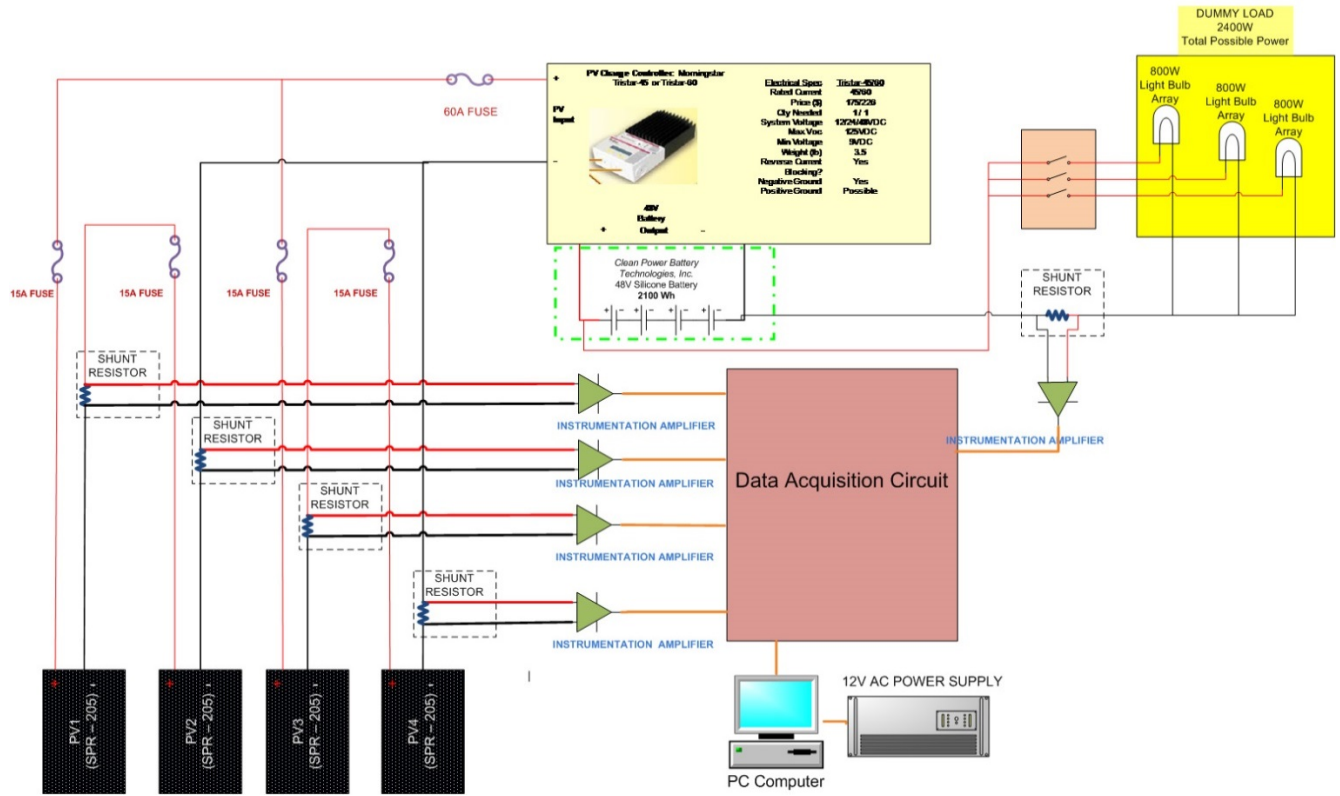

Fig. 2 Schematic of the Energy System Experiment

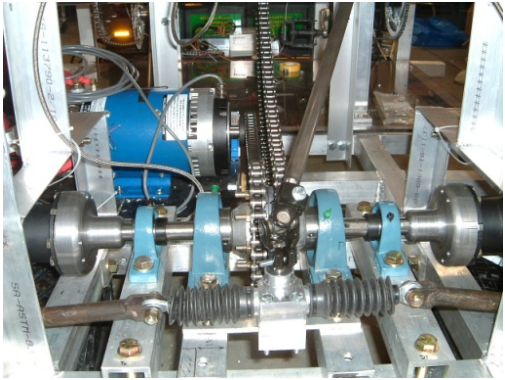

Fig. 3  The Power Transmission System

Friction clutches are integrated into the sprocket hubs mounted at the pedal input shaft which allows for direct power transmission from human pedaling at low speed of 5 to 7 mph. When the driving power of the vehicle is switched from human pedaling power to electric motor power, the clutches disengage and the pedal input shaft behaves like a free wheel. A hydraulic disc braking system is implemented in the front wheels of the ZEM vehicle.

## IV. DESIGN OF ZEM VEHICLE

The research on batteries and solar-electric power conversion and transmission as well as multiple power input to the drive train as mentioned above led to the design of complete power/drive system of the ZEM vehicle as illustrated in Figure 4.

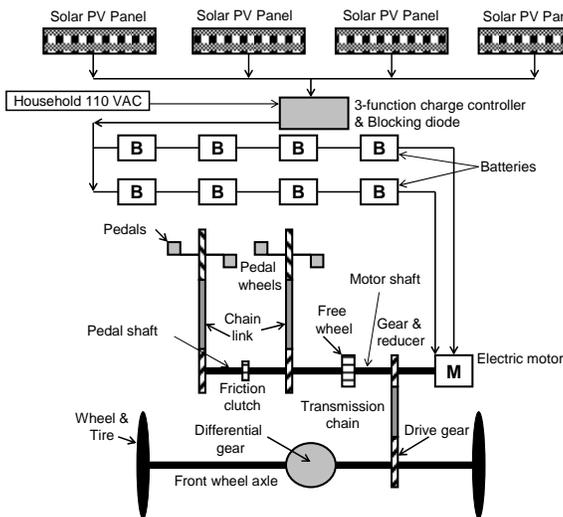

Fig. 4   Schematic of the Power Transmission System

Major components of the ZEM electrical system include a series motor, a motor controller unit, a silicone battery bank and charger, a solar PV system and its charge controller, a DC/DC conversion unit, and a microcontroller-based system to collect the data, process the information, and control the whole electrical power circuit.

The battery bank system includes eight 12V 70 Ahr silicone batteries, as shown in Figure 4. These 8 batteries are connected in series and parallel to provide 48 VDC power to the motor and DC/DC conversion unit. Measured results by the research work showed that for equivalent manufacturing specifications, the effectiveness ratios of silicone battery is higher than lead acid battery, but with lower charging rate. Since high energy density and environmental friendliness are more important to charging rate, silicone batteries were selected for the ZEM vehicle.

The solar PV system used in the project includes 4 silicon-based PV panels. These solar panels are rated with nominal voltage of 40V and current of 5.1A, with a peak power at 205W and conversion efficiency of 16.5%. Since the operating voltage input to the PV charger should be higher than 48V for better efficiency, the four PV panels are wired in two parallel strings of two panels in series. This connection provides the theoretical operating voltage of 80V with nominal current of about 10.2A and peak power at 410W. Experimental results showed that during the sunny periods, the system can reach a maximum operating voltage of about 56V, a maximum current of about 7A, and a peak power at about 350W. The solar PV power is used to charge the battery bank by a 3-function charge controller, which is rated at 45A from a 48-VDC input. This charge controller is able to share the load with the battery bank when the vehicle is in use and charge the battery whenever there is no load.

The motor used in the ZEM vehicle has maximum output of 10 hp operated with 48VDC supply. The velocity of the vehicle is controlled by a programmable solid state motor controller. This controller is rated at 48V and 500A and is controlled by 0-5kΩ potentiometer. The motor controller takes a signal from the potentiometer and varies the duty cycle of the pulse-width-modulated (PWM) voltage applied to the armature of the motor. Varying duty cycle of the PWM voltage effectively varies the average voltage applied to the motor. The motor controller is programmed to operate as desires via an RS-232 serial port.

The whole vehicle electrical system is controlled by a microcontroller based system and the data acquisition unit. The microcontroller, the data acquisition circuit, the dashboard display system, the vehicle lighting system, and all automatic relays and switches are all powered by 5 VDC and 12 VDC supplied by a 450W DC/DC conversion unit. In order to limit large current load as well as to save battery energy, the microcontroller system only turns on the motor when the forward speed of the vehicle exceeds 5 mph. This is not applied to vehicle reversal, and drivers are able to turn off this option if desired.

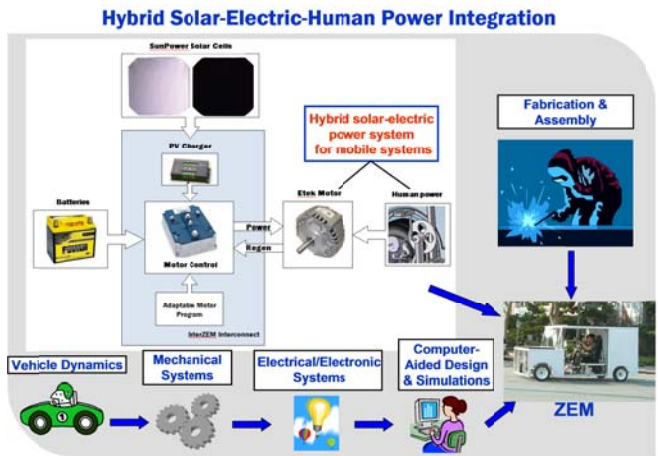

Fig. 5  Technologies for ZEM Vehicle Design

Figure 5 illustrates the synergistically integrated key technologies in the design of the ZEM vehicle. The design of the proof-of-concept prototype ZEM vehicle is shown in Figure 7, with 3 solar PV panels attached to the roof of the vehicle and another one on the top of the front hood. Eight high performance silicone batteries are placed under the seats.

The intimacy of mechanical and electronics systems for ZEM design required close dialogue between students of these two engineering disciplines. Consequently, in addition to encouraging one-on-one discussions among all student teams, there were expanded meetings scheduled twice during each semester with invited technical managers from local solar and EV industries, as well as senior officials from transportation department of the City of San Jose. These invited technical consultants offered valuable inputs and suggestions to all students in the project

## V. COMPLETION OF "PROOF-OF-CONCEPT" PROTOTYPE

A proof-of-concept prototype ZEM vehicle was constructed by 27 ME and 7 EE students and completed in the 2007-2008 academic year. They were organized into 6 functional sub-groups with approximately equal size memberships as shown in Figure 6.  Figure 7 shows the completed functional ZEM prototype.

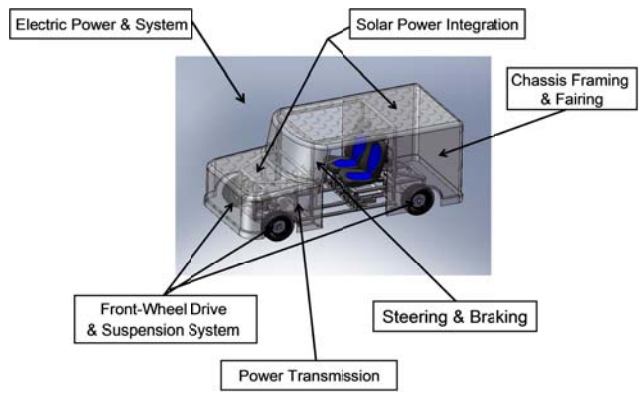

Fig. 6  Functional Grouping of Students in ZEM Prototype Construction

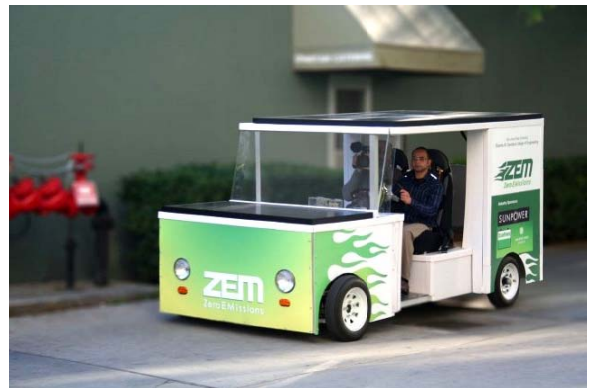

Fig. 7 The completed ZEM prototype

## VI. POSITIVE OUTCOMES OF THE ZEM PROJECT

The project not only provided the students with great learning experience in R&D, design, and fabrication, but also offered the students with unique opportunities in: (1) being mentored by Silicon Valley industry and business leaders in developing entrepreneurship, (2) learning strong leadership skills, (3) developing effective communication and organizational skills, and (4) developing fiscal responsibility with available funds and resources.  Based on our experience, the authors feel that the goal of the ZEM project is well accomplished.

### ACKNOWLEDGMENT

Authors are grateful for the continuous encouragements and financial supports by the College of Engineering at San Jose State University.

### REFERENCES

[1] "Shaping the Future-New Expectations for Undergraduate Education in Science, Mathematics, Engineering, and Technology," National Science Foundation, NSF-96-139, 1996.
[2] "Restructuring Engineering Education: A Focus on Change," National Science Foundation NSF 95-65, 1995.
[3] "Engineering Schools and Engineering Careers," American Society for Engineering Education, April 2009.
[4] "ABET 2001-2002 Criteria for Accrediting Engineering Programs," Accreditation Board for Engineering and Technology, 2002.
[5] McDonald, D., Devaprasad, J., Duesing, P, Mahajan, A., Qatu, M. and Walworth, M., "Re-Engineering the Senior Design Experience with Industry-sponsored Multidisciplinary Team Projects,"http://fie-conference.org/fie96/papers/398.pdf
[6] Ramachandran, R., Marchese, A.J., Ordonez, R., Sun, C, Constans, E., Schmalzel, J.L. and Newell, H.L. "Integration of Multidisciplinary Design and Technical Communication: An Inexorable Link," International Journal of Engineering Education, Vol. 18, No. 1, 2002, pp. 32-38.
[7] Burnell, L.J., Priest, J. W., and Durrett, J.R., "Assessment of a Resource Limited Process for Multidisciplinary Projects," Inroads-the SIGCSE Bulletin, Vol. 35, No. 4, December 2003.
[8] Guangdong Jiangmen Yuyang Special Batteries Company. LTD. Comparisons Between Lead-acid Batteries and GUINENG (silicon alloy) Power Batteries.  Retrieved February 11, 2008, from http://www.guineng.com/HTM/Jwork03.htm
[9] R. Swanson, M. Cudzinovic, D. DeCeuster, V. Desai, Jorn Jurgens, N. Kaminar, W. Mulligan, L. Rodrigugues-Barbarosa, D. Rose, D. Smith, A. Terao, and K. Wilson, "The Surface Polarization Effect of High Efficiency Silicon Solar Cells,"  15$^{th}$ PVSEC, Shanghai, China, 2005.

BIOGRAPHY

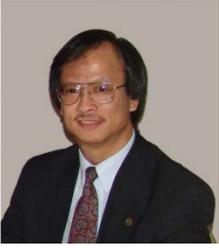

Raymond K. Yee is currently a Professor of the Department of Mechanical Engineering at San Jose State University, CA. Professor Yee received his BS degree with Highest Honors in Mechanical Engineering from Cal Poly, San Luis Obispo, MS degree and Ph.D. degree in Mechanical Engineering from the University of California at Berkeley. He is also a Registered Professional Mechanical Engineer in the State of California. Prior to joining San Jose State University, Professor Yee worked as a Member of the Technical Staff at AT&T Bell Labs in Naperville, Illinois and a senior consulting engineer in Silicon Valley. He has served as Associate Chair and Director of ME Program in the Department. His areas of interest in research and teaching include finite element method, mechanical design, mechanical behavior of materials, and fracture mechanics.

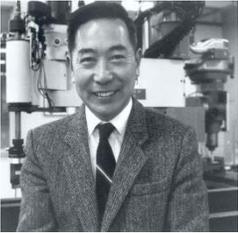

Tai-Ran Hsu, a Fellow of ASME, is currently a Professor and Chair of the Department of Mechanical Engineering at San Jose State University. He received a BS degree from National Cheng-Kung University in China; MS and Ph.D. degrees from University of New Brunswick and McGill University in Canada respectively. All his degrees were in mechanical engineering. He worked for steam power plant equipment and nuclear industries prior joining the academe. He taught mechanical engineering courses and served as department heads at universities in both Canada and USA. He has published over 120 technical papers in peer reviewed systems and eight books in finite element method in thermomechanics, computer-aided design, and a well-received textbook on microelectromechanical systems (MEMS) design and manufacture. The second edition of the latter book was published in March 2008.

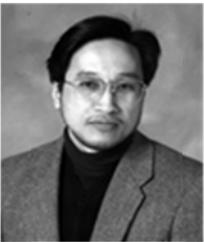

Thuy T. Le is currently a Professor and Associate Chair of Electrical Engineering Department at San José State University. In the past 20 years, he has held several research and teaching positions at U.S. national laboratories, universities, and private industries in the areas of SoC and ASIC design and verification, high-performance computing and computational nuclear reactor physics. He has served as General Chair, Technical Program Chair, and Technical Reviewer of many international conferences and technical journals, and published over 50 technical papers in peer reviewed systems. He earned his B.S., M.S., and Ph.D. degrees all from the University of California at Berkeley.